\documentstyle{amsppt}

\input epsf.def

\LimitsOnInts
\LimitsOnSums
\NoRunningHeads

\define \C {\Bbb C}
\define \R {\Bbb R}
\define \Z {\Bbb Z}
\define \Si {\Sigma}
\define \s {\sigma}
\define \la {\lambda}
\NoBlackBoxes

\topmatter
\title {Multiple Mellin-Barnes Integrals as Periods}\\ {of Calabi-Yau
Manifolds With Several Moduli}
\endtitle
\author{M.Passare, A.K.Tsikh, A.A.Cheshel} \endauthor
\abstract
We give a representation, in terms of iterated Mellin-Barnes integrals, of
periods on multi-moduli Calabi-Yau manifolds arising in superstring theory.
Using this
representation and the theory of multidimensional residues, we present a
method for analytic
continuation of the fundamental period in the form of Horn series.
\endabstract
\endtopmatter

Superstring vacua with $(2,2)$ worldsheet supersymmetry, and also
Landau-Ginz\-burg vacua, are
determined by the special geometry of the moduli spaces of Calabi-Yau
manifolds and of the
corresponding orbifolds. Moreover, the deformation moduli of the complex
structure and of
the K\"ahler metric constitute two sectors in the moduli space, and they
are in a $1-1$
correspondence with the $27$ and $\overline{27}$ multibraids of the matter
fields, which are
$E_6$-charged in the low energy limit of the theory.

Thanks to the mirror symmetry for Calabi-Yau manifolds the sector of
deformation moduli of
the K\"ahler metric is identified with the sector of deformation moduli of
the complex structure
on the corresponding mirror manifold [1]. The dynamics of the matter fields
in the low energy
theory depends on the geometry of the moduli space, and both the kinetic
term and the Yukawa
coupling constants are determined by a holomophic object on the manifold
$M$. This object is a
holomorphic $(3,0)$-form $\Omega$, or more precisely the period vector of
$\Omega$, having as
components
$$\varpi_j=\int\limits_{\gamma_j}\Omega, $$
where the $\gamma_j$ are basis cycles for the homology group. Mirror
symmetry allows one to study
the Yukawa field coupling constants as $(2,1)$-forms on the manifold $M$,
by calculating
the corresponding periods of $(1,1)$-forms on the mirror manifold $W$.

The idea of mirror symmetry in the theory of compactifications of
superstrings has been
intensively developed in the papers [1] - [11]. In particular, it has been
effectively used in
the study of periods on Calabi-Yau manifolds.

For a wide class of Calabi-Yau hypersurfaces (of arbitrary dimension $N-1$)
in weighted projective
space, the fundamental period of a holomorphic $(N-1)$-form $\Omega$ was
computed in the paper [2].
Considered as a function on the moduli space, this period is a
hypergeometric function, in fact
representable as a Horn series (cf.~Sect.1), which converges for large
values of the distinguished
modulus $\varphi_0$. Here arises the important problem to analytically
continue the period into
the region where the values of the modulus $\varphi_0$ are small, because
such an analytic
continuation allows one to calculate also the remaining periods of the form
$\Omega$. By means of an
artificial method for iterated summation of Horn series, the analytic
continuation was carried out in
[2], [3], [4] and [5], for some examples of hypersurfaces with two moduli.

In this paper we present a general method for analytic continuation of
periods that depend on two or
more moduli. Our approach consists in first representing the period as a
multiple Mellin-Barnes
integral (cf.~Sect.3). The Mellin-Barnes integral is then in its turn
calculated in terms of a
series of multidimensional Grothendieck residues, and this latter
computation is based on the
method of separating cycles, which was developed by one of the authors in
[13] and [14].

The idea of this method is based on the observation that the plane of
integration may be viewed as the
distinguished boundary (or edge) of many different polyhedra, and a choice
of any such polyhedron
that is also compatible with the form $\Omega$ produces its own residue
formula for the Mellin-Barnes
integral. (See [15] for the first realization of this idea in terms of a multidimensional Jordan
residue lemma.) In this way, the different residue formulas provide an
analytic continuation of the
Mellin-Barnes integral (MB), and hence of the original period. Our study of
the period may be
illustrated by the following 3-step scheme:
$$
\{ \text{ \it period } \} \rightarrow
\{ \text{ \it Horn series } \} \rightarrow
\{ \text{ \it MB integral } \} \rightarrow
\{ \text{ \it new Horn series } \}
$$

In carrying out the second and third steps of the scheme we use the simple
language of combinatorial
geometry, related to the mutual positions of the polar hyperplanes of those
gamma-functions that
occur in the construction of the Horn series and of the Mellin-Barnes integrals.

\head
{\bf 1. Periods on Calabi-Yau hypersurfaces in weighted projective spaces}
\endhead

A weighted $N$-dimensional projective space is determined by a choice of
$N+1$ natural numbers
(weights) $k=(k_0,k_1,\dots,k_N)$. Specifically, for non-zero vectors
$x=(x_0,x_1,\dots,x_N)$
in complex Euclidean space $\Bbb C^{N+1}$ one considers the equivalence
relation
$$x_i\sim\lambda^{k_i}x_{i},\phantom {hhhhh}   i=0,1,\dots,N ,$$
where $\lambda$ is any non-zero complex number. It is the quotient space
$\Bbb C^{N+1}/\sim$,
with respect to this equivalence relation, that constitutes the weighted
projective space of
dimension $N$ and weight $k_0,k_1,\dots,k_N$, and we denote it by ${\Bbb
P}_{(k_0,k_1,\dots,k_N)}^N={\Bbb P}_k^N$. A hypersurface in weighted
projective space ${\Bbb P}_k^N$
is defined to be the zero set
$$
M=\{x; P_0(x)=0\}
\tag{1.1}
$$
of a weighted-homogeneous polynomial (or superpotential) $P_0(x)$ of weight
$k$. This means that
$$P_0(x)=\sum_{\alpha\in
A}c_{\alpha}x^{\alpha}=\sum_{\alpha_0,\dots,\alpha_N}c_{\alpha_0,\dots,
\alpha_N}x_0^{\alpha_0}\cdots
x_N^{\alpha_N},$$
with every monomial $c_{\alpha}x^{\alpha}$ having the same weighted degree
$d=k_0\alpha_0+\dots
+k_N\alpha_N$. We shall be interested in smooth hypersurfaces, and these
are characterized the
condition that the polynomial $P_0$ and its gradient $dP_0$ should have no
common zeros in
$\Bbb C^{N+1}\setminus\{0\}$. Moreover, we are going to assume the condition
$$d=\sum_{i=0}^N k_i
\tag {1.2}
$$
to hold, that is, the degree of the hypersurface should coincide with the
sum of the components of
the weight $k$. Condition (1.2) guarantees that the hypersurface (1.1) is a
Calabi-Yau manifold:
The first Chern class of $M$ is trivial, and $M$ admits no non-trivial
holomorphic differential
forms of degree $p=1,2,\dots,N-2$. The next step in superstring theory is
to aconsider a deformation
of the manifold $M$ by means of a family of polynomials
$$
 P_{\varphi}(x)=P_0(x)-\varphi_0x_0x_1\dots x_N+\sum_{\beta\in B}\varphi_\beta
x^\beta.
$$
Here the standard monomial $x_0x_1\cdots x_N$ and the monomials
$x^{\beta},\beta\in B$,
that are added to $P_0(x)$, are chosen in such a way that their
coefficients $\varphi_0$ and
$\{\varphi_\beta\}$ provide a parametrization of the complex structures of
the manifold $M$. For
this reason these coefficients are called moduli, and the vector space
$\varphi=\{\varphi_0,\{\varphi_\beta\}\}$ is called the moduli space. In
the description of
the indicated complex structures, a crucial role is played by the
holomorphic differential $(N-1)$-form
$$\varOmega=\varOmega(\varphi)=Res_{M_\varphi}\left[\frac{\mu}{P_
\varphi(x)}\right] ,
\tag {1.3}
$$
where the residue $Res$ is taken with respect to the deformed submanifold
$$M_\varphi=\{x;P_\varphi(x)=0\} ,$$
along which the meromorphic form $\mu/P_\varphi(x)$ has a pole [12]. In
order that the quotient
$\mu/P_\varphi(x)$, which is written in homogeneous coordinates
$x=(x_0,x_1,\dots,x_N)\in\Bbb C^{N+1}$,
should be a well defined differential form on the weighted projective space
$\Bbb P_k^N$, in particular
invariant with respect to the equivalence $x_i\sim\lambda^{k_i}x_i$, it is
necessary that $\mu$ be
of the type
$$\mu=\sum_{i=0}^N(-1)^{i+1}k_ix_idx_0\dots\left[i\right]\dots dx_N ,
$$
where $\left[i\right]$ means that the index $i$ is deleted. Since the form $\varOmega$ is defined as
the residue of a meromorphic form, it follows from the Leray residue
formula ([12], Sect.21) that
each period of $\varOmega$ is given by some integral of the $N$-form
$\mu/P_\varphi(x)$ over a
$N$-dimensional cycle in $\Bbb P_k^N\setminus M_\varphi$. Any such integral
may in its turn be
reduced to a $(N+1)$-dimensional integral of the rational form
$dx/P_\varphi(x)$, with $dx=dx_0dx_1\dots
dx_N$, in the affine space $\Bbb C^{N+1}$, (cf.~[13], Sect.~20.1). In [16]
it was proved that, if the
denominator of a rational form $dx/Q(x)$ has fixed coefficients, then each
vertex of the Newton
polytope of $Q$, which does not lie on any of the coordinate planes,
determines a residue of the
rational form, and also that the collection of such residues is linearly
independent. In our case
some of the monomials in the denominator $P_\varphi(x)$ have variable
coefficients
$\varphi_\alpha$, and in this case one can associate a residue to every
monomial
$\varphi_\beta x^\beta $, for which $\beta$ is not in a coordinate plane,
by taking the chosen
coefficient $\varphi_\beta$ to be much bigger in absolute value than any
other of the
coefficients. Among the monomials with variable coefficients (which
contribute to the deformation of
the polynomial $P_0$) we have singled out the monomial $\varphi_0x_0\dots
x_N$, which in some sense
corresponds to the center of the Newton polytope for $P_0$. It turns out
that the residue associated to
this monomial may be computed by developing the integrand in a Laurent
series, and one calls this
residue a fundamental period for the form $\varOmega$. Its precise
definition is as follows [2]:
$$
        \omega_0(\varphi)=-\frac{\varphi_0}{(2 \pi i)^{N+1}}\int_\gamma
\frac{dx_0\cdots dx_N}{P_\varphi(x)},\phantom{h}|\varphi_o|\gg1,
\tag {1.4}
$$
the contour $\gamma$ being given as the Cartesian product of the $N+1$ circles
$|x_i|=1$, $i=0,1,\dots,N$.
The coefficient in front of the integral is chosen so as to simplify
notation in the
resulting formulas.

Let us label by $c_ix^{\alpha^i}$ the monomials in the fixed part $P_0$ of
the polynomial $P_\varphi$,
and by $\varphi_jx^{\beta^j}$ the monomials in the variable part (except
for $\varphi_0x_0\dots x_N$).
A computation (cf.~formula (2.21) in [2], where $c_i=1$) now yields
$$
        \varpi_0(\varphi)=\sum_{(n_i,m_j,r)\in
L} \frac{\Gamma(r+1)}
{\prod_i\Gamma(n_i+1)\prod_j\Gamma(m_j+1)}
\frac {\prod_ic_i^{n_i} \prod_j\varphi_j^{m_j} }
{\varphi_0^r} ,
\tag {1.5}
$$
where the summation is performed over the sublattice $L$, of the integer
lattice in the variables
$n_i$, $m_j$, $r$, defined by the following system of $N+2$ linear
homogeneous equations:
$$
\sum_i \alpha_k^in_i + \sum_j\beta_k^jm_j - r = 0,\quad k=0,\dots,N,
$$
$$
\sum_in_i + \sum_jm_j - r = 0 .
$$

Here $\alpha_k^i$ and $\beta_k^j$ are the coordinates of the vectors
$\alpha^i$, $\beta^j$ in
the lattice $\Bbb Z^{N+1}$. In the theory of multidimensional
hypergeometric functions Laurent
series such as (1.5) are called $\Gamma$-series (cf.~[17]). There exists a
standard procedure
to go from $\Gamma$-series to Horn series by chosing a basis for the
sublattice $L$. A Horn
series is the same as a power series with coefficients of a special kind:
They are quotients
of products of gamma-functions composed with linear functions of the
summation variables. In most
models of Calabi-Yau manifolds the number of monomials in the polynomial
$P_0$ is equal to
$N+1$, the number of variables. It follows that in these cases one can
incorporate the coefficients
$c_i$ into the variables, that is, one may assume $c_i=1$. The choice of a
basis for the sublattice
$L$ transforms the series (1.5) into a power series of Horn type
$$
\varpi_0(t)=\sum_{m\in \Bbb Z_+^n} \frac{\Gamma (\langle{}a,m\rangle{}+1)}
{\prod_{k=1}^q \Gamma (\langle{}a^k, m\rangle{}+1)} t_1^{m_1}\cdots t_n^{m_n},
\tag {1.6}
$$
where $n$ is the dimension of the moduli space (i.e. the number of coefficients
$\{\varphi_0, \varphi_j \}$), the vectors $a, a^k\in \Bbb R^n$ are
determined from the monomial
exponents $\alpha^i, \beta^j$, and the variables $t_i$ are monomials in the moduli
$\varphi_0, \varphi_j$.
Furthermore, the relation (weight balance)
$$
a= \sum_{k=1}^q a^k
\tag {1.7}
$$
is satisfied.

As an illustration of what has been said so far, and also to illustrate
results further on in
the paper, we consider the following example, borrowed from [2]. By means
of the polynomial
$$
P_0=x_0^7+x_1^7x_3+x_3^3+x_2^7x_4+x_4^3
\tag {1.8}
$$
we obtain a Calabi-Yau hypersurface of degree 21 in the weighted projective
space
$\Bbb P^4_{(3,2,2,7,7)}$. With this hypersurface there is associated a
2-parameter family of
manifolds
$$
M_\varphi =\{x_0^7+x_1^7x_3+x_3^3+x_2^7x_4+x_4^3-\varphi_0x_0x_1x_2x_3x_4+
\varphi_1x_0x_1^3x_2^6=0\},
$$
which characterizes the complex structures on the hypersurface $P_0=0$.
The fundamental period of the surface can be represented by a Horn series
(cf.~[2], formula (4.19))
$$
\sum_{m_1, m_2 \geq 0} \frac{\Gamma (7m_1+3m_2+1)}
{\Gamma^2(m_1+1)\Gamma^2(2m_1+m_2+1)m_1!m_2!} t_1^{m_1}t_2^{m_2},
\tag {1.9}
$$
where $t_1=\varphi_0^{-7}, t_2=\varphi_1/\varphi_0^3$, while the factorials
$m_i!=\Gamma(m_i+1)$
have been put aside to facilitate later computations.

\head
{\bf 2. Mellin-Barnes integrals and their representations as Horn series}
\endhead

By a $n$-dimensional Mellin-Barnes integral we mean an integral such as
$$
       \Phi (t)=\frac{1}{(2\pi i)^n}\int_{\gamma+i\Bbb
R^n}\frac{\prod_{j=1}^p\Gamma
(\langle{}a^j,z\rangle{}+b_j)}{\prod_{k=1}^q\Gamma(\langle{}c^k,z\rangle{}+d
_k)}(-t_1)^{-z_1}\cdots
(-t_n)^{-z_n}dz\phantom{h},
\tag{2.1}
$$
where $a^j$, $c^k$ are vectors in $\Bbb R^n$, the numbers $b_j$, $d_k$ are
also real,
and the vector $\gamma\in\Bbb R^n$ is chosen so that the real
$n$-dimensional set of
integration
$$\gamma+i\Bbb R^n=\{z\in\Bbb C^n;z_1=\gamma_1+iy_1,\dots,z_n=\gamma_n+iy_n\}
$$
does not intersect the polar hyperplanes
$$\langle{}a^j,z\rangle{}+b_j=-\nu,\phantom{hhh}\nu=0,1,2,\dots;\, j=1,\dots,p;
$$
finally, $dz=dz_1\cdots dz_n$.
In the one-variable case the integral (2.1) reduces to the following:
$$
        \frac{1}{2 \pi
i}\int_{\gamma-i\infty}^{\gamma+i\infty}\frac{\prod_j\Gamma(a_jz+b_
j)} {\prod_k\Gamma(c_kz+d_k)}(-t)^{-z}dz .
\tag{2.2}
$$
The integral (2.2) was studied by Barnes, and it represents the inverse
Mellin transform of the
integrand. The behaviour of the integrals (2.1) is to a large extent
governed by the following two
quantities. The first one is the vector
$$\Delta=\sum_ja^j - \sum_kc^k ,
\tag{2.3}
$$
and the second one is the scalar quantity
$$
        \alpha=\underset \parallel y \parallel =1 \to \min (\sum_j
|\langle{}a^j,y\rangle{}| - \sum_k
|\langle{}c^k,y\rangle{}|) ,
\tag{2.4}
$$
where $\parallel y \parallel = (y_1^2 + \dots + y_n^2)^{1/2}$ denotes the
Euclidean norm in
$\Bbb R^n $.
In the case $n=1$ we have
$$
        \alpha=\sum_j| a_j| - \sum_k| c_k| .
\tag{2.4a}
$$
The number $\alpha$ determines the values of the complex parameters
$t=(t_1,\dots,t_n)$,
for which the integral (2.1) converges, more precisely, it converges in the
domain (cf.~[15])
$$
         U = \{ t\in (\Bbb C \backslash 0)^n; \left[(\pi-\arg
t_1)^2+\dots+(\pi-\arg t_1)^2
\right]^{1/2} < \pi\alpha/2\},
$$
which is non-empty for $\alpha >0$. The quantity $\Delta$ characterizes the
domain in the
space of integration variables $z$, in which the integrand is a decreasing
function. The
asymptotic Stirling formula shows namely that, outside the polar set, the
expression under the
integral sign in (2.1) decreases exponentially at infinity in any proper
polyhedral cone in the
halfspace
$$
         \Pi_{\Delta ,\gamma}= \{z\in \Bbb C^n;\text{Re}
\langle{}\Delta,z\rangle{} < \langle{}\Delta,\gamma\rangle{}
\}.
$$
The vector $\Delta$ being real, this halfspace may be viewed as the direct
sum $\pi + i\Bbb R^n$
of the halfspace $\pi = \{ x\in \Bbb R^n;
\langle{}\Delta,x\rangle{} < \langle{}\Delta,\gamma\rangle{} \}$ in the
real subspace
$\Bbb R^n\subset \Bbb C^n$, and the imaginary subspace $i \Bbb R^n $. In
the the case $n=1$
the halfspace $\Pi_{\Delta ,\gamma}$ is just the left or right halfplane (depending on the sign
of $\Delta$), and $\pi$ is the left or right halfline on the real axis.
Hence the quantity
$\Delta$ plays a role similar to that of $\lambda$ in a Fourier type integral
$$
        \int_{-\infty}^{+\infty} f(x)e^{i\lambda x}dx ,\phantom{hh} \lambda
\in \Bbb R
,
\tag{2.5}
$$
which for suitable functions $f(x)$ may be computed by means of the well
known Jordan lemma
([18], Chap.5, Sect.2), as the sum of residues in the upper or lower
halfplane, depending on
the sign of $\lambda$. Now in the case where $\lambda=0$ this residue
computation may be performed
in either of the two halfplanes. We recall that the idea of the Jordan
lemma consists in the
following: The integral (2.5) is represented as the limit when $R\to\infty$
of the integrals
$\int_{-R}^{+R}$; the interval of integration $[-R,R]$ is completed to a
closed path by
adding a halfcircle in the upper or lower halfplane; to the obtained closed
contour one then
applies the Cauchy residue theorem. Concerning the one-variable integrals
(2.2), this idea was
carried out by Barnes and others (cf.~[19]), by replacing the upper and
lower halfplanes by the
left and right ones. And here it was the integrals (2.2) with zero
characteristic $\Delta$
that attracted the most interest, because then they represent all the
generalized hypergeometric
functions. For $\Delta=0$ the integral (2.2) as a sum of residues in the
left halfplane, as well
as in the right one. It then turns out that the sum of residues in the left
halfplane is a
Horn power series in the variable $t$, convergent in some disk $|t|<r$,
whereas the residue sum
in the right halfplane represent another Horn series in the variable
$t^{-1}$, which converges
outside the same disk; and if the quantity $\alpha$ is positive
(cf.~(2.4a)), then each of
these series is an analytic continuation of the other (cf.~[19]).

In the multidimensional case the situation is somewhat more complicated and
more varied. This is
because for $n=1$ the vertical axis of integration $\text{Re}\,z =\gamma$
is the boundary and
also the edge of precisely two polyhedral cones, namely the left and right
halfplanes.
By a polyhedral cone we understand a set of the form
$$\Pi = \{z\in \Bbb C^n; \text{Re}\, g_j(z)< r_j,\phantom{hh}j=1,\dots,m\},
$$
where $g_j(z)$ are linear functions with real coefficients. We shall be
interested in simplicial
cones, which correspond to $m=n$ with linearly independent functions $g_j$,
$j=1,\dots,n$ (such
cones have $n$ faces of maximal dimension). The coefficients being real,
one can write a
polyhedral simplicial cone as $\pi + i\Bbb R^n$, where $\pi$ is an octant
in the real subspace
$\Bbb R^n\subset \Bbb C^n$:
$$
\pi = \{x\in\Bbb R^n;\,g_j(x)<r_j,\, j=1,\dots,n \}.
$$
It follows that if $\gamma\in\Bbb R^n$ is the vertex of the octant $\pi$
(cf.~Fig.~1), then
the ``vertical" set of integration $\gamma+i\Bbb R^n$ in the integral (2.1)
is the edge of
the cone $\Pi$ (cf.~Fig.~2).

\midinsert
\centerline{\epsffile{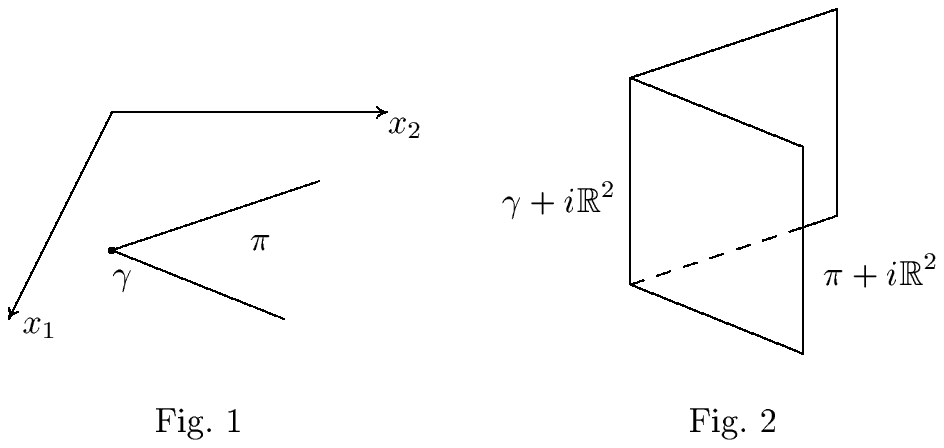}}
\botcaption{}\endcaption
\endinsert

\smallskip
Since there is an infinite number of possible octants that all have
$\gamma$ as their vertex,
it follows that the ``vertical" set of inegration is also the edge of
infinitely many
polyhedral cones. For the integral (2.1) one can accordingly obtain
different residue formulas
in cones with edge $\gamma +i\Bbb R^n$. Here we refer to local residues at
the points of intersection
of polar hyperplanes
$$
L_j^\nu =\{ \langle{}a^j,z\rangle{}+b_j=-\nu \}, \quad \nu=0,1,\dots, \quad
j=1,\dots,p.
\tag {2.6}
$$
At an intersection point
$$
z=z_J^m=L_{j_1}^{m_1}\cap\dots\cap L_{j_n}^{m_n},
$$
where precisely $n$ hyperplanes such as (2.6) come together, one may define
a local Grothendieck
residue
$$
\underset z_J^m \to {\text{res}}\, \omega =\frac{1}{(2\pi i)^n}
\int_{\gamma_J^m} \omega,
$$
where $\omega$ is the integrand from (2.1), and $\gamma_J^m$ is a
$n$-dimensional
cycle
$$
\gamma_J^m = \{z;|\langle{}a^{j_1},z\rangle{} + b_{j_1} + m| =\dots=
|\langle{}a^{j_n},z\rangle{} + b_{j_n}
+ m_n | = \varepsilon \},
$$
with $\varepsilon>0 $ sufficiently small. By means of the coordinate change
$$
        \text{w}_j = \langle{}a^j,z\rangle{} + b_j + m_j ,\quad j=1,\dots,n,
$$
and invoking the Cauchy formula, it is easily verified that
$$
\underset z_J^m \to {\text{res}}\, \omega =\frac{(-1)^{|m|}}{m!\,\Delta
_J}\frac{\prod_{j\ne
J}\Gamma(\langle{}a^j,z_j^m\rangle{}+b_j)}{\prod_{k=1}^q\Gamma(\langle{}c^k,
z_J^m\rangle{}+d_k)}
(-t_1)^{(z_J^m)_1}\cdots(-t_n)^{(z_J^m)_n} ,
\tag {2.7}
$$
where $\Delta_J=\det (a^{j_1},\dots,a^{j_n})$, $(z_J^m)_i$ are the
coordinates of the point
$z_J^m$, and $|m|=m_1+\dots+m_n$, $m!=m_1!\cdots m_n!$.

If there are more than $n$ complex hyperplanes that meet at the point $z$,
then this set of
hyperplanes is
divided into $n$ groups ($n$ divisors) $D_1,\dots,D_n$, and one defines the
Grothendieck residue
$\underset z \to {\text{res}}_{\{D_1,\dots,D_n\}}\,\omega $ with respect to
these divisors (cf.~[13],[20]).
In this case the residue formula becomes more complicated, and along with
powers of the variables
$t_j$ it will also involve powers of logarithms
$(\text{ln}\,t_1)^{p_1}\cdots (\text{ln}\,
t_n)^{p_n}$,
with $p_1+\dots+p_n$ at least $n$ units less than the number of hyperplanes
that intersect at the given point.
Now we notice that the expression (2.7) corresponds to a term in a Horn
series (its
coefficient is a quotient of products of gamma-functions with arguments
that are linear in
$m = (m_1,\dots,m_n)$). In the terminology of hypergeometric function
theory [17], the formula (2.7)
(i.e. the residue expression, when there are just $n$ intersecting polar
planes) corresponds to the
non-resonance case.

Next we observe that the solution $z_J^m$ to the system of linear equations
$$
\langle a^{j_1},z\rangle + b_{j_1} = -m_1,\ \dots\ \langle a^{j_n},z\rangle
+ b_{j_n} = -m_n
$$
can be linearly parametrized also with respect to $m=(m_1,\dots,m_n)$ as
follows:
$$
(z_J^m)_1 = \langle A^{j_1},m\rangle + B_{j_1},\ \dots\ (z_J^m)_n = \langle
A^{j_n},m\rangle + B_{j_n}.
$$
Hence, factoring out the monomial
$(-t)^{B_J}=(-t_1)^{B_{j_1}}\cdots(-t_n)^{B_{j_n}}$ in (2.7) and making the
substitution $\tau _i = t_1^{a_i^{j_1}}\cdots t_n^{a_i^{j_n}}$, $i =
1,\dots,n $, the expression (2.7) becomes a
term of a Horn power series in the variables $\tau_1,\dots,\tau_n$.

Now we turn to the question of deciding through which residues (2.7) one
should express
the integral (2.1). The case $n=2$, $\Delta \ne 0$ was considered in [15].
In this case, just like in
one variable, the integral (2.1) is represented by only one residue formula:
$$
\Phi (t) = \underset z_J^m\in \Pi_{\Delta,\gamma} \to {\sum} \underset z_J^m
\to {\text{res}}\, \omega,
$$
where the summation is over all intersection points $z_J^m =
L_{j_1}^{m_1} \cap L_{j_2}^{m_2}$ in the halfspace $\Pi_{\Delta,\gamma}$,
for which the lines
$L_{j_1}^{m_1}$ and $L_{j_2}^{m_2}$ intersect the line
$$
l = \{x \in \Bbb R^2; \langle{}\Delta ,x\rangle{} = \langle{}\Delta
,\gamma\rangle{}\}
$$
on different sides of the point $\gamma \in l $ (for this reason the cycle
of integration
$\gamma+i\Bbb R^2$ is said to be a separating cycle for the complex lines
$L_{j_1}^{m_1}$, $L_{j_2}^{m_2}$).

Turning to the case of interest to us $\Delta = 0 $,  we introduce the
following concept.
In the plane $\Bbb R^2$  we consider an arbitrary angle (cone) $\pi$ with
vertex $\gamma$
(cf.~Fig.~1). We shall say that the angle $\pi$ is compatible with the
family (2.7) of polar
lines $L_j^\nu $ of the integrand (2.1), if each such line intersects at
most one side of the
angle $\pi$ (in Fig.3 the angles $\pi_1, \pi_2, \pi_3$ are compatible with
the family
consisting of vertical, horizontal and oblique lines). Then in accordance
with the multi-dimensional
Jordan lemma [15], we have the following

\proclaim {Claim 1} Let $\Delta=0$. If the angle $\pi$ is compatible with
the family
$L_J^\nu =\{\langle{}a^j,z\rangle{} + b_j = - \nu \}$ of polar lines for
the integral (2.1),
then this integral is given by the residue formula
$$
\Phi (t) = \underset z_J^m \in \pi \to {\sum} \underset z_J^m \to
{\text{res}}\,\omega,
$$
where the summation is performed over all intersection points $z_J^m = L_{j_1}^{m_1}\cap
L_{j_2}^{m_2}$ in $\pi$, for which $L_{j_1}^{m_1}$ intersects only one side
of $\pi$, while
$L_{j_2}^{m_2}$ only meets the other side. Furthermore, the series
converges in a domain
of the form
$$
        \{t\in\Bbb C^2;\,|t_1^{A_1}t_2^{B_2}| <
r_1,\dots,|t_1^{A_s}t_2^{B_s}| < r_s \}.
$$
\endproclaim

\centerline{\epsffile{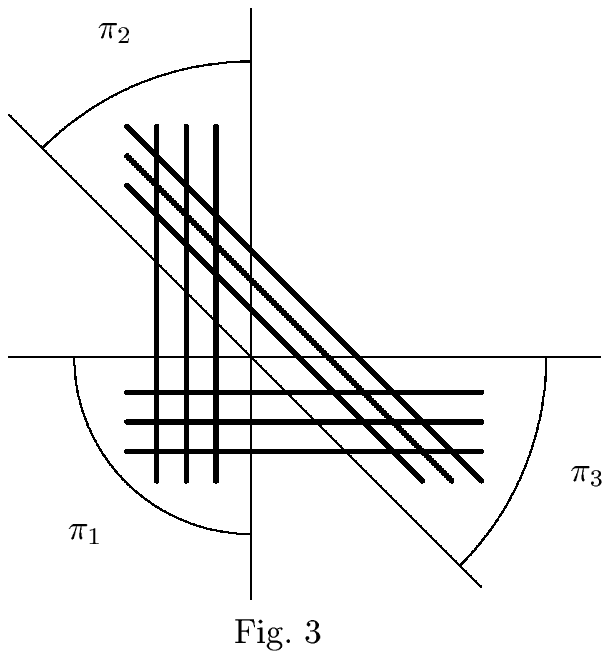}}

\smallskip

In sections 3 and 4 we will present the applications of the
multidimensional ($n>2$) Jordan lemma
that are of primary interest to us, but here we consider the following
two-dimensional example:
$$
\Phi (t) = \frac {1}{(2\pi i)^2} \int_{\gamma + i\Bbb R^2} \frac {\Gamma (z_1)
\Gamma (z_2) \Gamma (-7z_1-3z_2+1)}{\Gamma^2(-z_1+1)\Gamma^2(-2z_1-z_2+1)}
(-t_1)^{z_1}(-t_2)^{z_2}dz_1dz_2
\tag {2.8}
$$
>From the results in section 3 it will follow that this integral represents
the series (1.9), and hence
the fundamental period of the manifold (1.8). Here we have three families
of polar lines
$$L_1^\nu = \{z_1=-\nu \},\,L_2^\nu =\{z_2=-\nu\},\,L_3^\nu
=\{-7z_1-3z_2+1=-\nu \},
$$
as indicated in Fig.3. In the cone $\pi_1$ it is only lines from the first
two families that intersect,
i.e. poles of the gamma-functions $\Gamma(z_1)$ and $\Gamma(z_2)$. The
intersection points
$L_1^{m_1}\cap L_2^{m_2}$ can be parametrized as $(z_1,z_2) = (-m_1,-m_2)$,
and therefore, by
Claim 1 and formula (2.7), the integral $\Phi(t)$ is given by a Horn series
(1.9).

In this example $\pi_1$ is not the only angle that is compatible with the
polar family
$L_j^\nu $. Therefore there exist other residue formulas for $\Phi(t)$,
which also provide
analytic continuation of the series (1.9). Now, in $\pi_2$ it is the poles
of the functions
$\Gamma(z_1)$ and $\Gamma (-7z_1-3z_2+1)$ that intersect. These points of
intersection
admit the parametrization
$$
(z_1,z_2)=(-m_2,\frac 13m_1+\frac 73m_2+\frac 13).
$$
The determinant $\Delta_{31}$ equals $3$, and hence it follows from Claim 1
and formula
(2.7) that the integral (2.8) may be represented by the series
$$
\frac {(-t_2)^{-\frac 13}}{3} \underset m_1,m_2
\geqslant 0 \to {\sum }
\frac {\Gamma (\frac 13 m_1+\frac 73 m_2+1)\,
                                        e^{\pi i(\frac 23m_1-\frac 73m_2)}
\left(\frac {1}{t_2^{1/3}}\right)^{m_1}
\left(\frac {t_1}{t_2^{7/3}}\right)^{m_2}}
{m_1!m_2!\,\Gamma^2(m_2+1)\Gamma^2(-\frac 13 m_1-\frac 13 m_2+\frac 23)},
\tag {2.9}
$$
which is convergent in a domain in the variables $t_1$, $t_2$, which is
different from
the one for the series (1.9).

Finally, in the angle $\pi_3$ it is the poles of the functions
$\Gamma(z_2)$ and $\Gamma
(-7z_1-3z_2+1)$ that meet. The intersection point are given by
$$
(z_1,z_2) = (\frac 37m_1+ \frac 17m_2+ \frac 17, -m_1).
$$
The determinant $\Delta_{23}$ is equal to $7$, and so this time we obtain
the representation
$$
\frac {(-t_1)^{-\frac 17}}{7} \underset m_1,m_2 \geqslant 0 \to {\sum }
\frac {\Gamma (\frac 37m_1+ \frac 17m_2+ \frac 17)\,
                  e^{\pi i(\frac 67m_2-\frac 37m_1)}
\left(\frac {t_2}{t_1^{3/7}}\right)^{m_1}
 \left(\frac 1{t_1^{1/7}}\right)^{m_2}}
{m_1!m_2!\,\Gamma^2(-\frac 37m_1 -\frac 17m_2+ \frac 67) \Gamma^2
(\frac 17m_1-\frac 27m_2+\frac 57)}.
\tag {2.10}
$$
The fact that the series (2.9) and (2.10) are analytic continuations of the
series (1.9) is
a consequence of a result that we provide in section 4.

\head
{\bf 3. Representation of periods by Mellin-Barnes integrals}
\endhead

We know that the fundamental period $\varpi$ can be expressed as a Horn
series (1.6).
We shall now represent this series in the form of a Mellin-Barnes integral
(2.1).

\proclaim {Claim 2} If in the series (1.6) the vector $a$ have positive
components,
then this series admits a representation as the Mellin-Barnes integral
$$
\varpi _0(t) = \frac 1{(2\pi i)^n} \int _{\gamma + i\Bbb R^n} \frac {\prod
_{j = 1}^n \Gamma (z_j) \Gamma (1-z_j) \Gamma (1 -
\langle{}a,z\rangle{})}{\prod _{k=1}^q
\Gamma (1-\langle{}a^k,z\rangle{})}(-t)^{-z}dz,
\tag {3.1}
$$
where $\gamma $ is any point in the polytope
$$\Pi =\{x \in \Bbb R^n;\,0<x_j<1,\,j=1,\dots,n,\,\langle{}a,x\rangle{}<1\,\},$$
and $(-t)^z=(-t_1)^{-z_1}\cdots(-t_n)^{-z_n}$.
\endproclaim

To prove this claim we consider the polyhedron $\Pi = \pi + i\Bbb R^n $, where
$$
\pi = \{x \in \Bbb R^n;\,x_j \leqslant \gamma _j,\,j=1,\dots,n\},
$$
with $\gamma_j$ being the components of the vector $\gamma$. The integrand
in (3.1) has $2n+1$
families of polar hyperplanes
$\{z_j=-\nu\}$, $\{1-z_j=-\nu \}$, and $\{1-\langle{}a,z\rangle{}=-\nu\}$,
of which only the first $n$ families intersect the polyhedron $\Pi$. Each
of these families form a
divisor
$$ D_j = \bigcup_{\nu =0}^{\infty}\{z_j = -\nu \}, \quad j=1,\dots,n.$$
The collection of divisors $D_j$ is compatible with the polyhedron $\Pi$
(cf.~[15]) in the sense that
each $D_j$ is disjoint from the corresponding face of $\Pi$. More precisely, if
$\pi_j =\{x \in \pi ; x_j=\gamma_j\}$ is a face of the octant $\pi$, then
$\Pi_j =\pi_j + i\Bbb R^n$
is a face of the polyhedron $\Pi$, and the compatibility conditions
$$
D_j\cap \Pi_j = \varnothing,\quad j=1,\dots,n
$$
are clearly satisfied.
By the abstract multidimensional Jordan lemma [15] the integral (3.1) is
equal to the sum of
Grothendieck residues of the integrand in the polyhedron $\Pi$. Each
residue is located at
a point $(-m_1,\dots, -m_n)$, and it is equal to
$$
\multline
\frac {(-1)^{| m | }}{m!} \frac {\prod _{j=1}^{n}\Gamma (1+m_j) \Gamma
(1+\langle{}a,m\rangle{})}{\prod_{k=1}^q\Gamma
(1+\langle{}a^k,m\rangle{})}(-t)^m \\
=\frac {\Gamma (1+\langle{}a,m\rangle{})}{\prod _{k=1}^q\Gamma
(1+\langle{}a^k,m\rangle{})}\,t_1^{m_1},
\dots,t_n^{m_n}.
\endmultline
$$
This completes the proof of our claim.

\head
{\bf 4. Analytic continuation of the periods}
\endhead

We observe that we have already carried out the first two steps in our
scheme for the study of the
fundamental period, that we mentioned in the introduction of this paper.
Indeed, in section 1 we
went through the first step: \{{\it period}\} $\Longrightarrow $ \{{\it
Horn series}\} (formula (1.6)),
and in section 3 the second step: \{{\it Horn series}\} $\Longrightarrow $
\{{\it
Mellin-Barnes integral}\} was accomplished (formula (3.1.)). The results of
section 3 will now allow
us to perform the third and final step in the scheme, i.e. to calculate the
Mellin-Barnes integrals in
various ways, thereby establishing the analytic continuation of the
fundamental period.

Let us assume that in the series (1.6) the vector $a$ has positive
components $a_j$, and that all
factors $\Gamma (m_j+1)=m_j!$, $j=1,\dots,n$, are present in the
denominator of the coefficients.
Let these be the last factors $\Gamma (m_j + 1) = m_j!$, $j=1,\dots,n$,
where $s=q-n$. It is not hard
to see that in this case the integral representation (3.1) of the series
will have the appearence
$$
\varpi_0(t) = \frac 1{(2\pi i)^n} \int _{\gamma + i\Bbb R^n} \frac {\prod
_{j=1}^n
\Gamma (z_j)\, \Gamma (1- \langle{}a,z\rangle{})} {\prod _{k=1}^s \Gamma (1
- \langle{}a^k,z\rangle{})}\,(-t)^{-z}\,dz,
\tag {3.2}
$$
with the balance condition
$$
a - \sum_{k=1}^s a^k = (1,\dots,1)
$$
for the weights.
{\bf Remark.}\ The authors are not aware of any case
of a series (1.6) representing a period on a Calabi-Yau hypersurface in
weighted projective space,
for which the positivity condition on the coordinate vector $a$ is not
satisfied.
However, in the paper [2] (formula (5.3)) an example of a period is given
for which the factorials
$m_i!$ are missing in its series representation of type (1.6). It is
conspicuous that the corresponding Calabi-Yau manifold lacks a mirror
partner in weighted projective space [2], [8].
As shown in the paper [11], this missing mirror image may however be
realized as a hypersurface in
a more general toric manifold [9].
\smallskip

We now formulate our main result regarding the integral (4.1):

\proclaim {Claim 3} For each $j$ from 1 to $n$ the integral (4.1) admits
the series
representation
$$
\varpi_j(t) = \frac {(-1)^{j-1}}{a_j} \sum _{m\ge0} \frac
{(-1)^{|m|}\Gamma\left(
\frac{1+\langle
A^j,m\rangle}{a_j}\right)(-t_1)^{m_1}\cdots(-t_j)^{-\frac{1+\langle A^j,m
\rangle}{a_j}}
\cdots(-t_n)^{m_n}}{m!\prod_{k=1}^s\Gamma\left((a_j-a_j^k+\langle
A^{kj},m\rangle)/a_j\right)},
$$
where $a_j$, $a_j^k$ are the components of the vectors $a$ and $a^k$, the
vector $A^j$ equals
$(a_1,\dots,1,\dots,a_n)$ with 1 in the $j$'th slot, the vectors $A^{kj}$
are given by
$$(a_ja_1^k-a_1a_j^k,\dots,-a_j^k,\dots,a_ja_n^k-a_na_j^k),$$
and as usual $|m|=m_1+\dots+m_n$ and $m!=m_1!\cdots m_n!$. Moreover, if for
the integral (4.1)
the quantity $\alpha$, defined by formula (2.4), is positive, then each of
the series $\varpi_j(t)$
provides an analytic continuation of the fundamental period $\varpi_0(t)$
to its corresponding domain
of convergence.
\endproclaim

For instance, for the integral (2.8) the quantity $\alpha$ is equal to
$2/\sqrt{58}$, and hence
the series (2.9), (2.10) do give analytic continuations of the series (1.9).

To prove the formulated claim one may argue as in the proof of Claim~2.
Indeed, for each
$j=1,\dots,n$ we consider the polyhedron $\Pi_j=\pi_j+i\Bbb R^n$, where
$\pi_j$ is the octant
$$\pi_j=\{x\in\Bbb R^n; x_i\le\gamma_i, i=1,\dots[j]\dots,n,\ \langle
a,x\rangle>\langle a,\gamma\rangle
\}.$$
It is evident that $\gamma+i\Bbb R^n$ is an edge for this polyhedron, and
that the polar divisor of
the function $\Gamma(z_j)$ does not intersect $\Pi_j$. Furthermore, the divisors
$$D_1=\{z_1=-m_1\},\dots, D_j=\{1-\langle a_j,z\rangle=-m_j\},\dots,
D_n=\{z_n=-m_n\},$$
which represent the poles of the functions $\Gamma(z_1),\dots,
\Gamma(1-\langle a_j,z\rangle),
\dots, \Gamma(z_n)$, are compatible with the polyhedron $\Pi_j$. Hence,
according to the multidimensional
Jordan lemma the integral (4.1) is equal to the sum of Grothendieck
residues with respect to the
divisors $D_1,\dots, D_n$ at their intersection points in $\Pi_j$. Applying
the residue formula
at these points, we find that the integral is equal to $\varpi_j(t)$ as claimed.

If the quantity $\alpha$, given in formula (2.4), is positive for the
integral (4.1), then the
domain of convergence for this integral will intersect each of the
respective domains of convergence for the
series $\varpi_j(t)$. Consequently, in this case these latter series
analytically continue
$\varpi_0(t)$ into their respective convergence domains.

In conclusion we remark that one-dimensional Mellin-Barnes integrals  have
been commonly used in
mathematical physics [21] in connection with questions of asymptotics of
determinants in string
theory, high temperature limits for free energy of
scalar fields in curved space-time etc. It is our belief that it would be
adequate to study some of
these questions in greater generality by means of multidimensional
Mellin-Barnes integrals.

{\bf Acknowledgements} Passare and Tsikh are indebted to the Royal Swedish
Academy of Sciences for
its financial support. Cheshel wishes to express his gratitude towards V.
Rabotin and B. Rubtsov
for useful discussions, and towards A. Niemi for his hospitality at the
Department of Theoretical
Physics of Uppsala University (Sweden).

\newpage

\end